\documentclass[aps,prd,twocolumn,superscriptaddress,altaffilletter,nobibnotes,nofootinbib,10pt,showpacs,showkeys,preprintnumbers]{revtex4-2}

\usepackage{float}
\usepackage{graphicx}
\usepackage{dcolumn}
\usepackage[dvipsnames]{xcolor}
\usepackage{bm}
\usepackage{lineno}
\usepackage{amsmath}
\usepackage{soul}
\RequirePackage{multirow}

\usepackage{tikz,xcolor,hyperref}

\definecolor{lime}{HTML}{A6CE39}
\DeclareRobustCommand{\orcidicon}{
	\begin{tikzpicture}
	\draw[lime, fill=lime] (0,0) 
	circle [radius=0.16] 
	node[white] {{\fontfamily{qag}\selectfont \tiny ID}};
	\draw[white, fill=white] (-0.0625,0.095) 
	circle [radius=0.007];
	\end{tikzpicture}
	\hspace{-2mm}
}
\foreach \x in {A, ..., Z}{\expandafter\xdef\csname orcid\x\endcsname{\noexpand\href{https://orcid.org/\csname orcidauthor\x\endcsname}
			{\noexpand\orcidicon}}
}

\def\ims{0.45}
\usepackage{subcaption}

\begin{document}
\title{Investigation of Nuclear Modification Factor from RHIC to LHC energies using Boltzmann Transport equation in conjunction with $q$-Weibull distribution}

\author{Rohit Gupta\orcidA{}}
 \email{rohitg876@gmail.com}
\affiliation{Department of Physics, Kashi Naresh Government Post Graduate College, Gyanpur, Bhadohi 221304, Uttar Pradesh, India}

\begin{abstract}
The study of nuclear modification factor is crucial in advancing our knowledge of the hot and dense nuclear matter created during high energy heavy-ion collision. In this direction, we have developed a theoretical model for the nuclear modification factor using the Boltzmann Transport equation in relaxation time approximation with the $q$-Weibull distribution as the final state distribution and studied the experimental data of nuclear modification factor of charged hadrons as well as identified particles at various energies ranging from $7.7$ GeV measured at RHIC upto the maximum value of $5.44$ TeV studied in LHC. We observed a good agreement between the model and the experimental data as can be quantified using the $\chi^2/NDF$ values. We have also studied the mass dependence of different fit parameters that appears in the theoretical model and observe a linear mass dependence of some parameters.
\end{abstract}
\maketitle
\section{Introduction}
Understanding the deconfined quark matter \cite{Shuryak:1980tp} created during very early stage of high energy heavy-ion collision is among one of the core motivation behind ongoing and upcoming particle physics experiments. Several experiments collide heavy-ions with center of mass energies ranging from as low as $\sqrt{s_{NN}} = 7.7$ GeV at RHIC all the way upto $5.44$ TeV per nucleon pair at LHC with a quest to study the dynamics of Quark Gluon Plasma (QGP) created during such heavy-ion collision. Since this QGP state is being created at extremely high energy density and for very short interval of time, the technological limitations makes it impossible to directly probe the QGP state. However, there are certain signatures that signals the formation of this state in heavy-ion collision. Some of the signatures include $J/\psi$ suppression \cite{CMS:2012bms}, strangeness enhancement \cite{Koch:1986ud, ALICE:2013xmt}, jet quenching \cite{CMS:2011iwn} and collective flow \cite{ALICE:2011ab}  etc.

Some of the above signatures such as collective flow \cite{CMS:2016fnw} and strangeness enhancement \cite{ALICE:2016fzo}  have recently been discovered in high multiplicity $pp$ collision, however, one of the signature that is so far unique to heavy-ion collision is jet quenching. In this article, our primary focus is on study of jet quenching. Jets are a clustered beam of partons, all moving in approximately same direction. 

Jet quenching \cite{Wang:1992qdg} refers to loss of energy of partons due to gluon emission as they pass through the strongly interacting QGP medium. This energy loss leads to suppression in the production of high $p_T$ particles which are produced from parton fragmentation. The variable which is used to quantify the jet quenching is called the Nuclear Modification factor $(R_{AA})$. As the name suggests, $R_{AA}$ measures the impact of the nuclear environment on particle production by comparing yields in nucleus-nucleus collisions to those in proton-proton collisions, normalized for the number of binary collisions. It is quantified in terms of the ratio of yields as \cite{ALICE:2012aqc}:
\begin{equation}
    R_{AA}(p_T) = \frac{d^2N_{AA}/dydp_T}{\langle T_{AA} \rangle d^2 \sigma_{pp}/dydp_T}
\end{equation}
Here, $T_{AA}$ is the nuclear overlap function and is given
as $\langle T_{AA} \rangle = \langle N_{coll}\rangle /\sigma^{NN}_{inel}$ where $\sigma^{NN}_{inel}$ is the inelastic nucleon-nucleon cross section and $\langle N_{coll}\rangle$ represent average number of binary nucleon-nucleon collision.
 In the above equation, $N_{AA}$ is the particle yield in nucleus-nucleus (AA) collision and $\sigma_{pp}$ is the particle yield in $pp$ collision.  Another variant of the nuclear modification factor is commonly used and is given in terms of ratio of yields in peripheral and central heavy-ion collision as \cite{STAR:2017ieb}:
\begin{equation}
    R_{CP}(p_T) = \frac{\langle N_{coll}\rangle_{Peripheral}}{\langle N_{coll}\rangle_{Central}}\frac{(d^2N/d\eta dp_T)_{Central}}{(d^2N/d\eta dp_T)_{Peripheral}}
\end{equation}
Here, $C$, $P$ refers to central and peripheral collision respectively. In the above equation for $R_{AA}$ and $R_{CP}$, if we consider the heavy-ion collision as a incoherent superposition of $N_{coll}$ independent $pp$ collision then we should get the value of $1$ for entire $p_T$ range. A value greater than $1$ represent enhancement whereas a value less than $1$ represent suppression in particle production in heavy-ion collision compared to $pp$ collision.
\begin{figure}[!h]
\centering
  \includegraphics[width=\ims\textwidth]{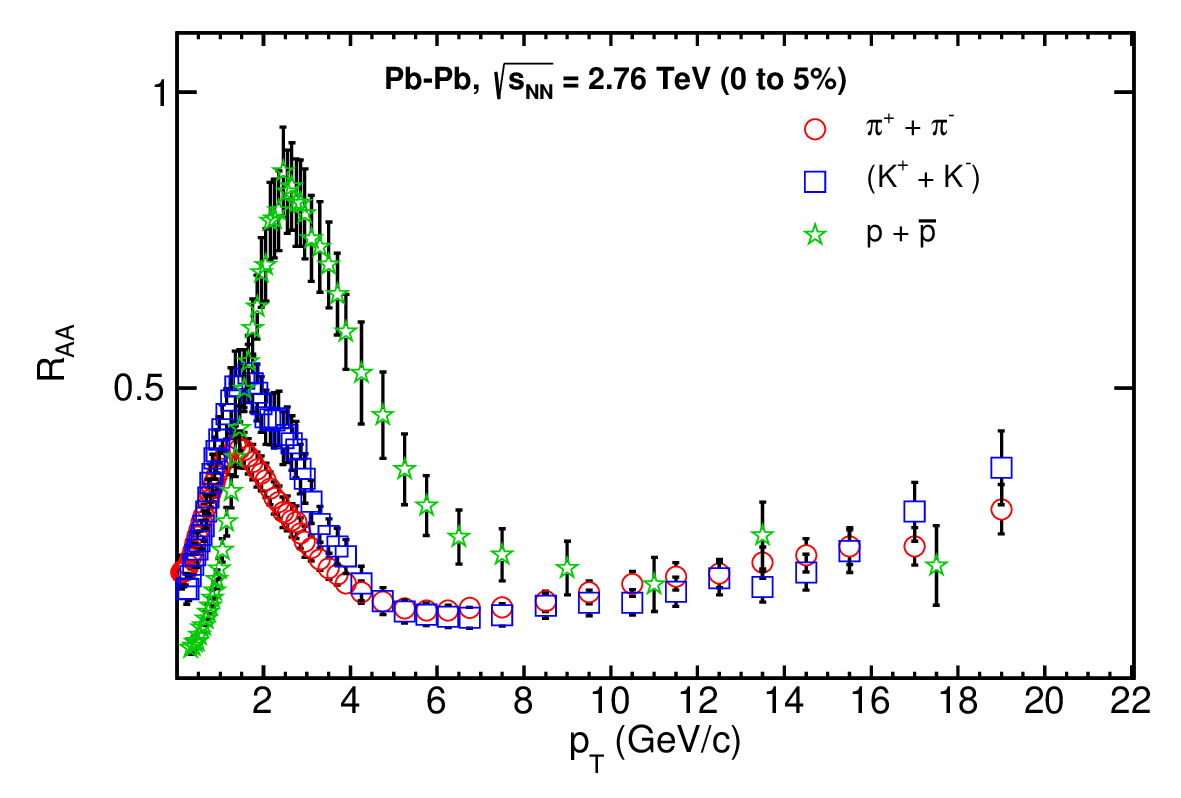}
  \caption{Nuclear modification factor of identified hadrons produced in most central $PbPb$ collision at $2.76$ TeV \cite{ALICE:2014juv}.}
  \label{fig:RAA_identified_data}
  \end{figure}
  
  The experimental data for $R_{AA}$ of identified hadrons produced in $2.76$ TeV $PbPb$ collision in given in Fig.~\ref{fig:RAA_identified_data} for charged pions, kaons and protons. The trend in the pattern of nuclear modification factor provides us with the crucial information about the system created during heavy-ion collision. A suppression in the high $p_T$ region points toward the formation of strongly interacting QGP medium whereas an enhancement in the intermediate $p_T$ region is linked with Cronin enhancement \cite{Cronin:1974zm, Accardi:2002ik, STAR:2017ieb}.

Nuclear modification factor has been studied across varied energies, collision system and centralities for both identified particles and charged hadrons. ALICE experiment has measured the $R_{AA}$ for identified particles such as charged pions, kaons, protons and $K^*(892)^0$, $\phi(1020)$, $J/\psi$ mesons at $2.76$ TeV \cite{ALICE:2014juv, ALICE:2017ban, ALICE:2013osk} and $5.02$ TeV \cite{ALICE:2019hno, ALICE:2021ptz, ALICE:2016flj}. 
The ALICE experiment has also measured the $R_{AA}$ for the combined yield of primary charged particles at a center of mass energy of $2.76$ TeV \cite{ALICE:2012aqc}, $5.02$ TeV \cite{ALICE:2018vuu} and $5.44$ TeV \cite{ALICE:2018hza}. Similar measurement at RHIC energies has been performed for charged hadrons by PHENIX experiment at $130$ GeV \cite{PHENIX:2002diz} and STAR experiment at $200$  \cite{STAR:2003fka}, $62.4$, $39$, $27$, $19.6$, $14.5$, $11.5$ $\&$ $7.7$ GeV \cite{STAR:2017ieb}.

As is evident from the discussion above, nuclear modification factor depends upon the transverse momentum $(p_T)$ spectra of particles and hence it is customary to discuss phenomenological models of transverse momentum spectra in order to have a better understanding of pattern in the nuclear modification factor. Although the standard choice to study the energy or momentum distribution of final state particles that are produced in heavy-ion collision is Quantum Chromodynamics (QCD) which is the underlying theory to study the partonic interactions. The constraints put up by the high value of coupling strength at low energies makes it difficult to apply perturbative QCD theory to study the transverse momentum spectra. In this context, several phenomenological models have been developed with different physics inputs \cite{Schnedermann:1993ws, Stodolsky:1995ds, Tsallis:1987eu, Jena:2020wno, Gupta:2020naz, Ristea:2013ara, Tang:2008ud, Gupta:2021efj, Dash:2018qln} to study the $p_T$ spectra. These models include Boltzmann-Gibbs statistics \cite{Schnedermann:1993ws, Stodolsky:1995ds}, Tsallis statistics \cite{Tsallis:1987eu}, Blast-Wave (BW) and Tsallis Blast-Wave (TBW) \cite{Ristea:2013ara, Tang:2008ud} etc. These models are widely used to study the spectra, however, their applicability is limited to low $p_T$ range upto $p_T$ values of $3-4$ GeV/c \cite{Gupta:2021efj}. In the high $p_T$ regime, where hard scattering processes dominate particle production, these phenomenological models often deviate from the experimental data. Since in experimental data for nuclear modification factor, suppression is observed beyond $2-3$ GeV/c in majority of the case so it is necessary to choose a phenomenological model that can fit the $p_T$ data for broader $p_T$ range beyond $3-4$ GeV/c. Hence, we have chosen the $q$-Weibull distribution function \cite{Dash:2018qln} which is shown to nicely fit the experimental data of transverse momentum spectra and multiplicity distribution for a wider range \cite{Dash:2018qln, Nayak:2018avf, Sharma:2018vsy, Behera:2016hak, Gupta:2023ggq}. 

Boltzmann transport equation (BTE) in relaxation time approximation (RTA) has been used extensively to study different aspects of high energy collision including the study of anisotropic flow \cite{Akhil:2023xpb}, temperature fluctuation \cite{Bhattacharyya:2015nwa} and nuclear modification factor $(R_{AA})$ \cite{Tripathy:2017kwb, Tripathy:2016hlg, Singh:2023agf, Qiao:2020yry, Moriggi:2022xbg}. In the study of $R_{AA}$ using BTE, the choice of final state and equilibrium distribution function has always been a point of discussion. In this context, we have used the $q$-Weibull distribution as the final state distribution of particles. For the equilibrium distribution we have considered the Boltzmann-Gibbs distribution which is the most natural choice for a thermally equilibrated system.

The detailed calculation of Boltzmann transport equation within relaxation time approximation is provided in the next section followed by the results of the fit obtained using this formalism to the experimental data of nuclear modification factor. In the end we will provide a conclusion of the work discussed in this manuscript.

\section{Theoretical description of Nuclear Modification factor using BTE in RTA}
Let us consider a distribution function $f(r,p,t)$ representing the distribution of particles in a phase space at position $(r)$, momentum $(p)$ and time $(t)$ such that $\int f(r,p,t)d^3rd^3p = N$, the total number of molecules in a given phase space. From kinetic theory of gases, considering the collision and other molecular interactions, the equation of motion for the distribution function can be given in terms of Boltzmann transport equation as \cite{huang:1987}:
\begin{equation}
    \frac{df(r,p,t)}{dt} = \frac{\partial f}{\partial t} + \textbf{v}.\nabla_rf + \textbf{F}.\nabla_p f = C[f]
\end{equation}
where $\textbf{v}$ is the velocity and $\textbf{F}$ is the external force on the molecule. In the above equation $\nabla_r$, $\nabla_p$ represent the gradient operator with respect to $r$ and $p$ respectively. Before proceeding further into solving the above differential equation, it is crucial to specify the collision term $C[f]$. Considering a homogeneous system $(\nabla_rf = 0)$ which is devoid of any external force $(\textbf{F} = 0)$, the above equation simplifies to,
\begin{equation}
    \frac{df(r,p,t)}{dt} = \frac{\partial f}{\partial t}  = C[f] 
     \label{simple_BTE}
\end{equation}
The collision term in BTE can be estimated for a system close to equilibrium using the relaxation time approximation with BGK-type collision term \cite{Bhatnagar:1954zz} which assumes that the collision tend to exponentially drive system toward equilibrium \cite{Kumar:2017bja, Bhadury:2020ngq, Baym:1984np}. In this formalism the collision term is given as,
\begin{equation}
    C[f] = -\frac{f-f_{eq}}{\tau}
    \label{coll_term}
\end{equation} 
where $\tau$ represent the relaxation time (the time taken by a non-equilibrium system to reach equilibrium) and $f_{eq}$ represent the Boltzmann equilibrium distribution. Putting above form of collision term Eq.~(\ref{coll_term}) into the simplified BTE Eq.~(\ref{simple_BTE}) we get:
\begin{equation}
    \frac{\partial f}{\partial t} = -\frac{f-f_{eq}}{\tau}
\end{equation}
with its solution as:
\begin{equation}
    f_{fin} = f_{eq} + (f_{in} - f_{eq})e^{-\frac{t_f}{\tau}}
    \label{final_dist}
\end{equation} 
In order to arrive at the above equation we have used the following initial conditions, at $t=0$, $f = f_{in}$ and at $t=t_f$, $f=f_{fin}$ where $t_f$ is the freeze-out time, $f_{in}$ is initial state distribution and $f_{fin}$ is final state distribution. Rearranging the terms in above equation we get:
\begin{equation}
    f_{in} = f_{eq} + (f_{fin}-f_{eq})e^{\frac{t_f}{\tau}}
        \label{initial_dist}
\end{equation}
Specifying $R_{AA}$ as a ratio of final to initial distribution of particles and using Eq.~(\ref{initial_dist}) we get:
\begin{equation}
    R_{AA} = \frac{f_{fin}}{f_{in}} = \frac{f_{fin}}{f_{eq} + (f_{fin}-f_{eq})e^{\frac{t_f}{\tau}}}
\end{equation}
We can also write this equation in inverse form as:
\begin{equation}
   R_{AA} =  \left[ \frac{f_{eq}}{f_{fin}} + \left( 1-\frac{f_{eq}}{f_{fin}} \right)e^{\frac{t_f}{\tau}}    \right]^{-1}
   \label{RAA_initial}
\end{equation}
This is the master equation utilizing the BTE within the framework of RTA to estimate the nuclear modification factor. 

It is important to point out here that this approach is fundamentally different from the other works in the direction of utilizing the BTE to study the nuclear modification factor. Existing works \cite{Tripathy:2017kwb, Tripathy:2016hlg, Singh:2023agf, Qiao:2020yry, Moriggi:2022xbg} have utilized the master equation for $R_{AA}$ in terms of $f_{eq}$ $\&$ $f_{in}$ instead of $f_{eq}$ $\&$ $f_{fin}$ as in the Eq.~\ref{RAA_initial}. In order to approximate the functional form of $f_{eq}$ $\&$ $f_{in}$, different distribution functions such as Boltzmann distribution, Tsallis distribution, Boltzmann Gibbs Blast-Wave (BGBW) and Tsallis Blast-Wave (TBW) models etc. have been considered. Since, we do not have any information about the initial distribution $f_{in}$ and on the contrary the final distribution $f_{fin}$ (the transverse momentum spectra of final state particles that are free streaming to the detector after the freeze-out) have been studied extensively across broad range of energies, multiplicities and collision type using different distribution functions \cite{Schnedermann:1993ws, Stodolsky:1995ds, Tsallis:1987eu, Jena:2020wno, Gupta:2020naz, Ristea:2013ara, Tang:2008ud, Gupta:2021efj, Dash:2018qln}. It appears more natural to consider the equation of $R_{AA}$ in terms of $f_{fin}$ as in the Eq.~\ref{RAA_initial} instead of $f_{in}$ and then utilize the best option among several phenomenological models for the distribution of final state $p_T$ spectra.

In order to utilize the Eq.~\ref{RAA_initial} to study the experimental data, next task is to specify the functional form of distribution functions $f_{eq}$ and $f_{fin}$. 
For a system of particles in thermal equilibrium, most natural choice for the distribution function is the Boltzmann-Gibbs (BG) distribution. So, we have chosen the BG distribution function as our equilibrium function $f_{eq}$ given as:
\begin{equation}
    f_{eq} = \frac{1}{2\pi p_T}\frac{d^2N}{dp_Tdy} = \frac{gV}{(2\pi)^3}m_T~ exp\left(\frac{-m_T}{T}\right)
    \label{boltz}
\end{equation}
In above equation, $g$ is the spin degeneracy factor, $m_T$ is transverse mass given as $m_T = \sqrt{m^2 + p_T^2}$ and $V$ is the system volume. 

   \begin{figure*}
       \centering
\begin{subfigure}[b]{0.45\textwidth}
\includegraphics[width=\textwidth]{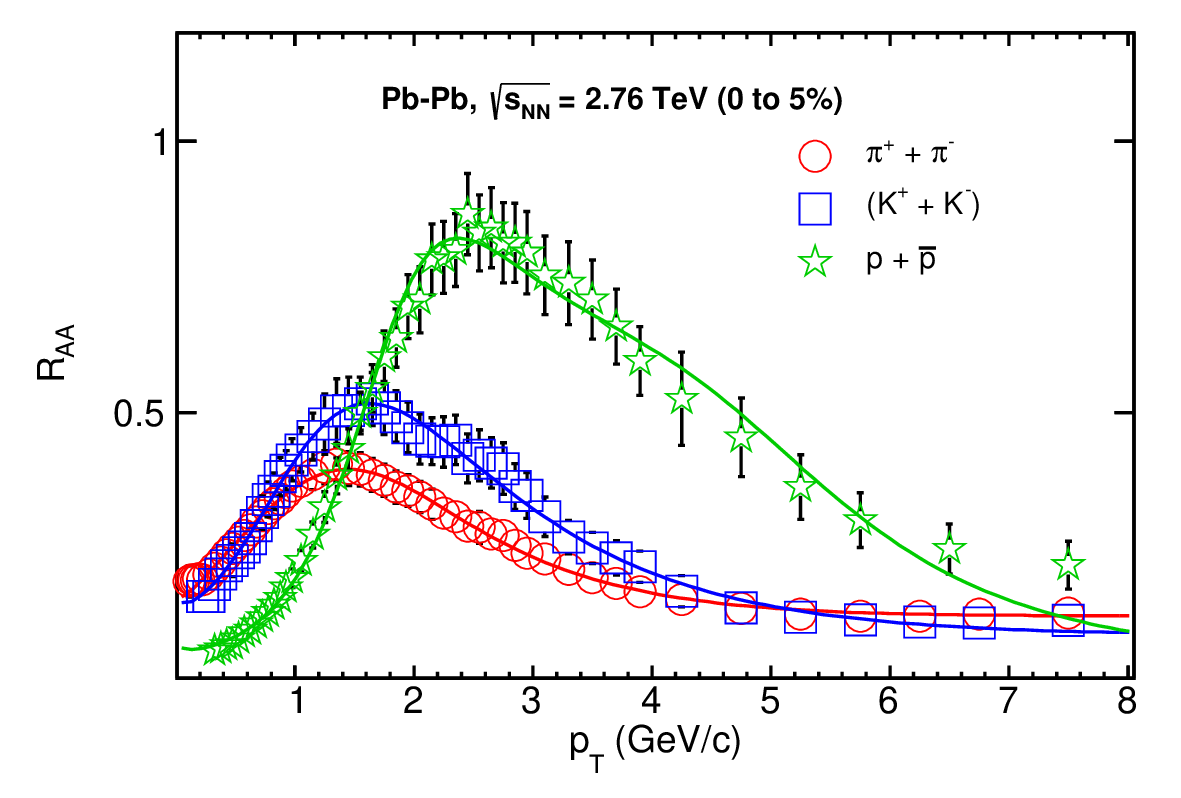}
\caption{$R_{AA}$ of pions, kaons and protons at $2.76$ TeV \cite{ALICE:2014juv}}
\label{fig:2760_1}
\end{subfigure}
\begin{subfigure}[b]{0.45\textwidth}
\includegraphics[width=\textwidth]{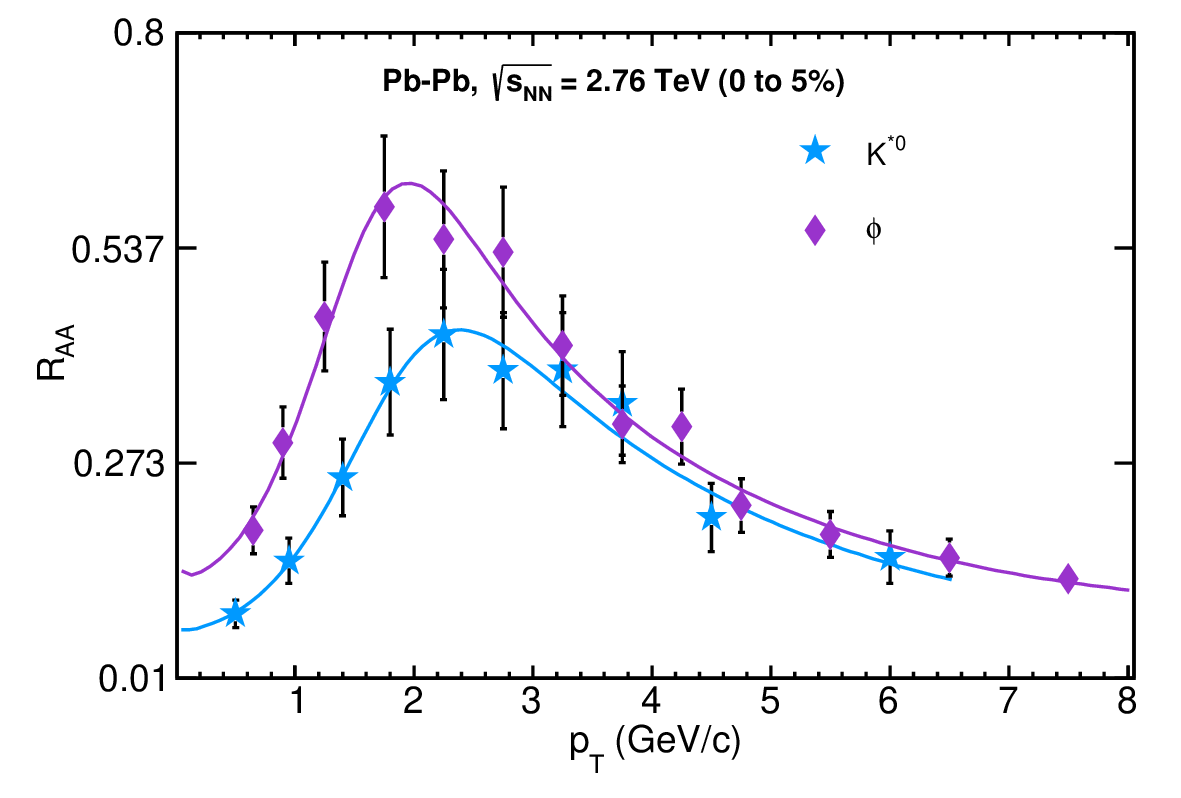}
\caption{$R_{AA}$ of $K^*(892)^0$ and phi meson at $2.76$ TeV \cite{ALICE:2017ban}}
\label{fig:2760_2}
\end{subfigure}
\begin{subfigure}[b]{0.45\textwidth}
\includegraphics[width=\textwidth]{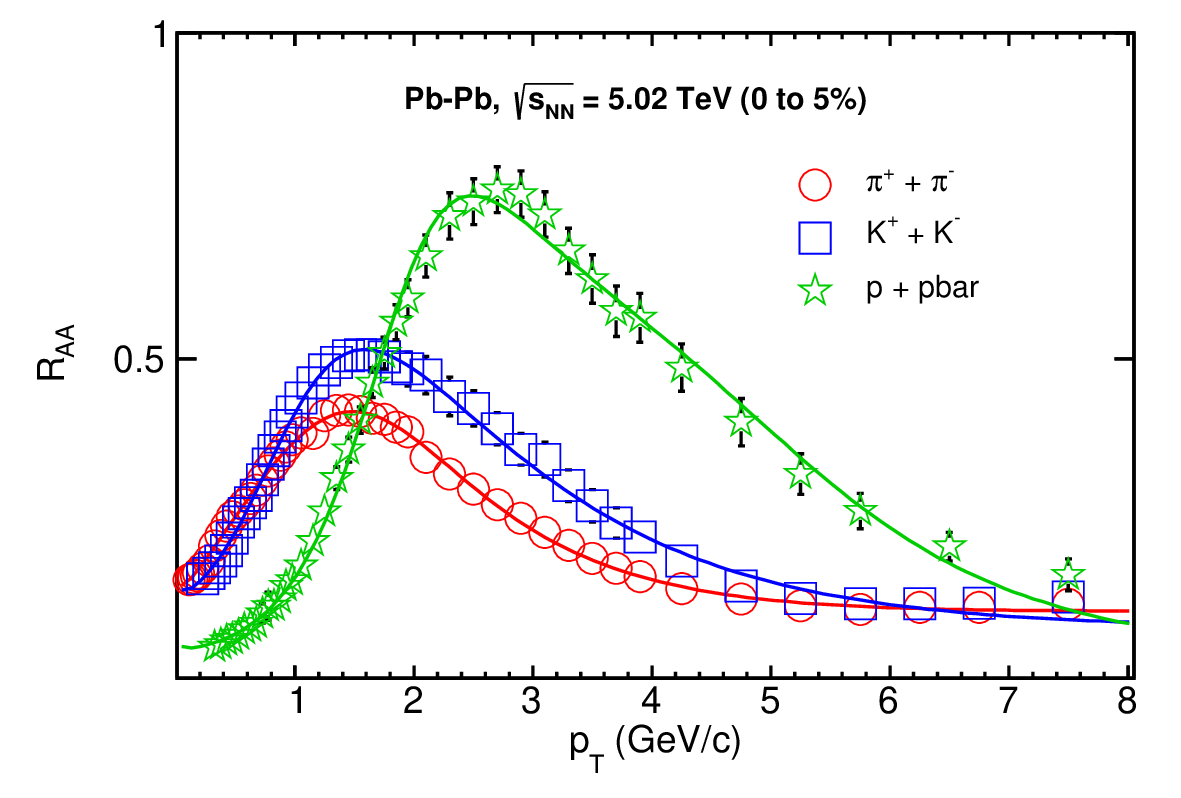}
\caption{$R_{AA}$ of pions, kaons and protons at $5.02$ TeV \cite{ALICE:2019hno}}
\label{fig:5020_1}
\end{subfigure}
\begin{subfigure}[b]{0.45\textwidth}
\includegraphics[width=\textwidth]{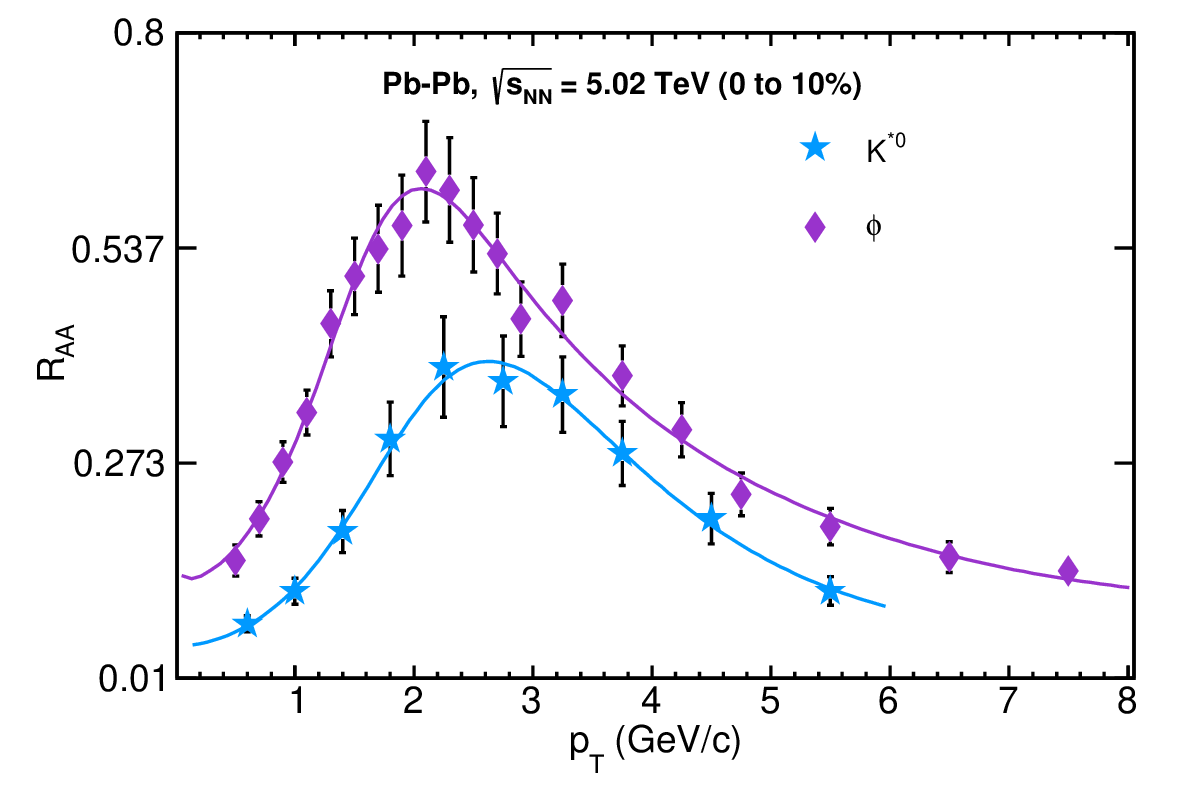}
\caption{$R_{AA}$ of $K^*(892)^0$ and phi meson at $2.76$ TeV \cite{ALICE:2021ptz}}
\label{fig:5020_2}
\end{subfigure}
\caption{Nuclear modification factor of different identified particles produced in most central $PbPb$ collision at $2.76$ $\&$ $5.02$ TeV fitted with the equation for $R_{AA}$ given in Eq.~\ref{RAA_final}}
\label{fig:identified}
    \end{figure*}
    
Since, $f_{fin}$ represent the distribution at the freeze-out time $t_f$ after which particles will free stream to the detectors, so, as already discussed in the introduction section, best choice to describe this final state distribution of particles over a broader $p_T$ range is the $q$-Weibull distribution function. This distribution function is an extension of Weibull distribution developed in 1951 by a Swedish mathematician Waloddi Weibull. The probability distribution function in case of Weibull distribution is given as:
\[
P(x,\lambda,k)=
\begin{cases}
\frac{k}{\lambda} \left( \frac{x}{\lambda}\right)^{k-1}\; e^{-(x/\lambda)^k} & x \geq 0\\
0 & x<0
\end{cases}
\]
where $\lambda$ represent the scale parameter and $k$ is the shape parameter and the values of $k$ $\&$ $\lambda$ is greater than zero. In several works \cite{brown1989, brown1995derivation}, Weibull distribution has been used successfully to describe the process in which the dynamical evolution of the system is driven by the fragmentation and sequential branching. The Tsallis generalization of the Weibull distribution was first proposed in the Ref. \cite{Dash:2018qln} and the generalized probability distribution called $q$-Weibull distribution function is given as:
\begin{equation}
P_q(x,q,\lambda,k)=\frac{k}{\lambda} \left( \frac{x}{\lambda}\right)^{k-1}\; e_q^{-(\frac{x}{\lambda})^k}
\label{qweibull_eq}
\end{equation}
where 
\begin{equation}
e_q^{-(\frac{x}{\lambda})^k} = \left( 1-(1-q)\left( \frac{x}{\lambda}\right)^k\right)^{\left( \frac{1}{1-q} \right)}   
\end{equation}
In the above equation, parameter $k$ is linked to the onset of hard scattering process and parameter $\lambda$ is related to the collective expansion velocity of hadrons \cite{Dash:2018qln}. Also, parameter $q$ is connected to the degree of non-extensivity in a thermodynamical system.
The $q$-Weibull distribution function is being used extensively to study the transverse momentum spectra of final state particles produced in high energy collision across different energies and multiplicities. So, it will be interesting to consider this as our final state distribution function $f_{fin}$ and check how well the final equation for nuclear modification factor matches with the experimental data. Using $q$-Weibull, final distribution function will of the form:
\begin{equation}
    \begin{aligned}
        f_{fin} = \frac{1}{2\pi p_T}\frac{d^2N}{dp_Tdy} = & D\frac{k}{\lambda} \left( \frac{p_T}{\lambda}\right)^{k-1}\\
        &\left( 1-(1-q)\left( \frac{p_T}{\lambda}\right)^k\right)^{\left( \frac{1}{1-q} \right)}  
    \end{aligned}
    \label{f_final}
\end{equation}
where $D$ is just a normalization parameter.
Putting Eq.~\ref{f_final} as our final distribution $f_{fin}$ and Eq.~\ref{boltz} as equilibrium distribution $f_{eq}$ in master equation for $R_{AA}$ Eq.~\ref{RAA_initial}, we get:

\begin{equation}
\begin{aligned}
     R_{AA} = \biggr[& \frac{\frac{gV}{(2\pi)^3}m_T~ exp\left(\frac{-m_T}{T}\right)}{D\frac{k}{\lambda} \left( \frac{p_T}{\lambda}\right)^{k-1}\; e_q^{-(\frac{p_T}{\lambda})^k}} +\\
         &\left(1-\frac{\frac{gV}{(2\pi)^3}m_T~ exp\left(\frac{-m_T}{T}\right)}{D\frac{k}{\lambda} \left( \frac{p_T}{\lambda}\right)^{k-1}\; e_q^{-(\frac{p_T}{\lambda})^k}} \right) e^{\frac{t_f}{\tau}}\biggr]^{-1}
\end{aligned}
\end{equation}
For the purpose of simplification, we have replaced $\frac{gV}{D(2\pi)^3}$ with a single normalization parameter $D'$. With this modification, final equation that is used for the purpose of fitting the experimental data is given as:
\begin{equation}
\begin{aligned}
     R_{AA} = \Bigg[& \frac{m_T~ exp\left(\frac{-m_T}{T}\right)}{D'\frac{k}{\lambda} \left( \frac{p_T}{\lambda}\right)^{k-1}\; e_q^{-(p_T/\lambda)^k}} +\\
         &\left(1-\frac{m_T~ exp\left(\frac{-m_T}{T}\right)}{D'\frac{k}{\lambda} \left( \frac{p_T}{\lambda}\right)^{k-1}\; e_q^{-(p_T/\lambda)^k}} \right) e^{\frac{t_f}{\tau}}\Bigg]^{-1}
\end{aligned}
\label{RAA_final}
\end{equation}
We have used this equation to fit the experimental data of nuclear modification factor of different species produced in collision of heavy-ions such as $AuAu$, $PbPb$ $\&$ $XeXe$ over a broad range of energies. The results obtained by fitting Eq.~\ref{RAA_final} with the experimental data will be discussed in the next section.

\begin{table*}[t]
\centering
\caption{Values of fit parameters obtained by fitting the experimental data of nuclear modification factor of particles produced in heavy-ion collision at different energies with the Eq.~\ref{RAA_final} obtained using the BTE}
\label{table_1}
\begin{tabular*}{\textwidth}{@{\extracolsep{\fill}}ccccccc@{}}
\hline
\multicolumn{1}{@{}l}{Particle type} & Energy (GeV) & $k$ &$\lambda$&$q$&$t_f/\tau$&$\chi^2/NDF$\\ 
\hline
 \multirow{10}{*}{Charged Hadrons} &$5440$&$1.103460$&$0.221443$&1.007287&$1.860654$&$0.157153$ \\ \cline{2-7}
&$5020$&$1.145497$&$0.240551$&$1.010394$&$2.148230$&$0.222213$ \\ \cline{2-7}
&$2760$&$1.149288$&$0.237893$&$1.012644$&$2.087201$&$0.026645$ \\ \cline{2-7}
&$200$&$1.048003$&$0.196725$&$1.002941$&$1.755912$&$0.124791$ \\ \cline{2-7}
&$130$&$1.099313$&$0.226323$&$1.003405$&$2.784152$&$0.293685$ \\ \cline{2-7}
&$62.4$&$1.066601$&$0.207253$&$1.003354$&$1.207884$&$0.262297$ \\ \cline{2-7}
&$27$&$1.085351$&$0.216412$&$1.003177$&$1.220898$&$0.179805$ \\ \cline{2-7}
&$19.6$&$1.059206$&$0.203384$&$1.001285$&$1.165683$&$0.085581$ \\ \cline{2-7}
&$11.5$&$0.986209$&$0.167179$&$0.996946$&$0.986306$&$0.097113$ \\ \cline{2-7}
&$7.7$&$1.001434$&$0.175754$&$0.997578$&$1.139056$&$0.105112$ \\ \hline
\multirow{1}{*}{$\pi^++\pi^-$}&\multirow{5}{*}{5020}&$1.138138$&$0.233499$&$1.013911$&$2.177862$&$0.397461$ \\ \cline{1-1} \cline{3-7}
\multirow{1}{*}{$K^++K^-$}&&$1.189292$&$0.278436$&$1.009732$&$2.408234$&$0.473149$ \\ \cline{1-1} \cline{3-7}
\multirow{1}{*}{$p+\overline{p}$}&&$1.240824$&$0.331346$&$1.007214$&$3.291584$&$0.693280$ \\ \cline{1-1} \cline{3-7}
\multirow{1}{*}{$K^{*0}$}&&$1.229703$&$0.316954$&$1.008415$&$3.310143$&$0.123356$ \\ \cline{1-1} \cline{3-7}
\multirow{1}{*}{$\phi$}&&$1.247083$&$0.339090$&$1.008140$&$2.393895$&$0.383204$ \\ \cline{1-1} \cline{3-7}
\multirow{1}{*}{$J/\psi$}&&$1.581026$&$0.752138$&$1.013214$&$1.335297$&$0.597107$ \\
\hline
\multirow{1}{*}{$\pi^++\pi^-$}&\multirow{5}{*}{2760}&$1.124854$&$0.227401$&$1.013529$&$2.077124$&$0.083898$ \\ \cline{1-1} \cline{3-7}
\multirow{1}{*}{$K^++K^-$}&&$1.199294$&$0.280606$&$1.011560$&$2.393015$&$0.150296$ \\ \cline{1-1} \cline{3-7}
\multirow{1}{*}{$p+\overline{p}$}&&$1.253442$&$0.339334$&$1.007744$&$3.259797$&$0.533376$ \\ \cline{1-1} \cline{3-7}
\multirow{1}{*}{$K^{*0}$}&&$1.188240$&$0.296546$&$1.005548$&$2.914350$&$0.275011$ \\ \cline{1-1} \cline{3-7}
\multirow{1}{*}{$\phi$}&&$1.252478$&$0.343484$&$1.008344$&$2.394666$&$0.310883$ \\ \cline{1-1} \cline{3-7}
\multirow{1}{*}{$J/\psi$}&&$1.533541$&$0.753457$&$1.008932$&$1.354616$&$0.544854$ \\
 \hline
 \\
\end{tabular*}
\end{table*}
\begin{figure}[!h]
\centering
  \includegraphics[width=\ims\textwidth]{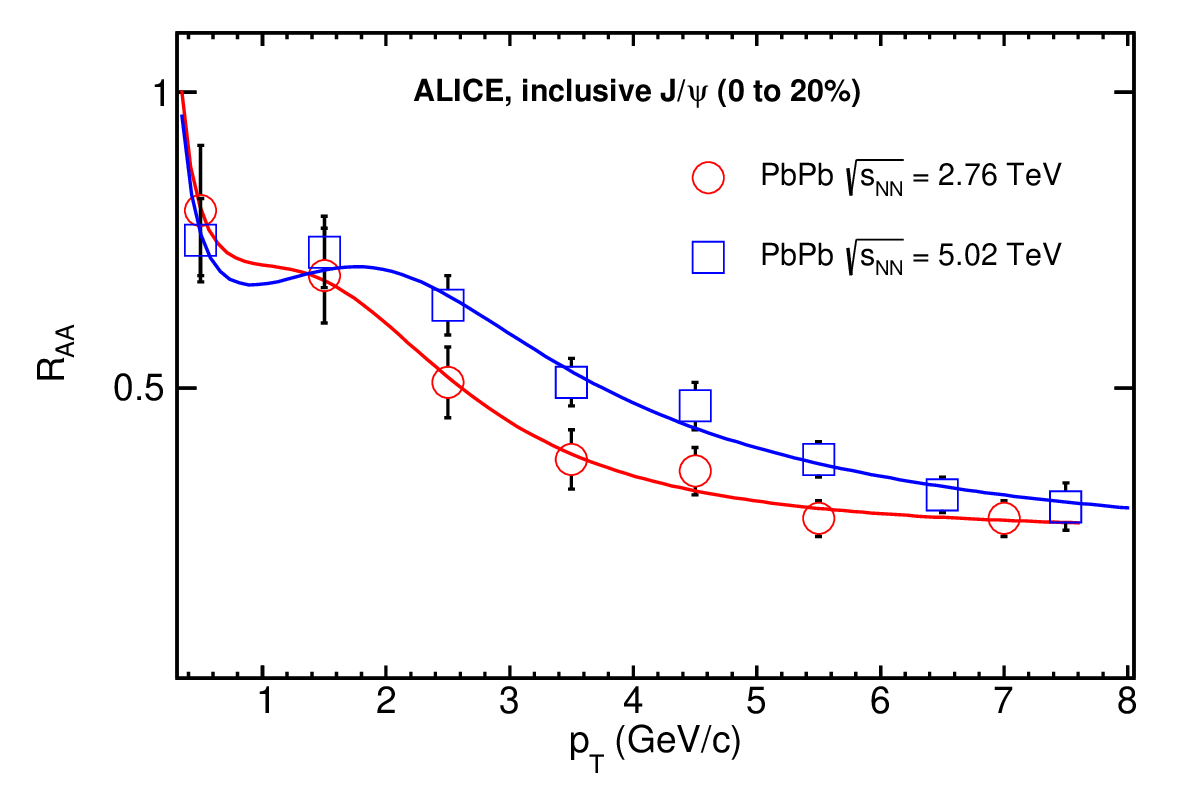}
  \caption{Nuclear modification factor of inclusive $J/\psi$ produced in most central $PbPb$ collision at $5.02~\&~2.76$ TeV fitted with the equation for $R_{AA}$ given in Eq.~\ref{RAA_final}.}
  \label{fig:RAA_jpsi}
  \end{figure}
\begin{figure}[!h]
\centering
  \includegraphics[width=\ims\textwidth]{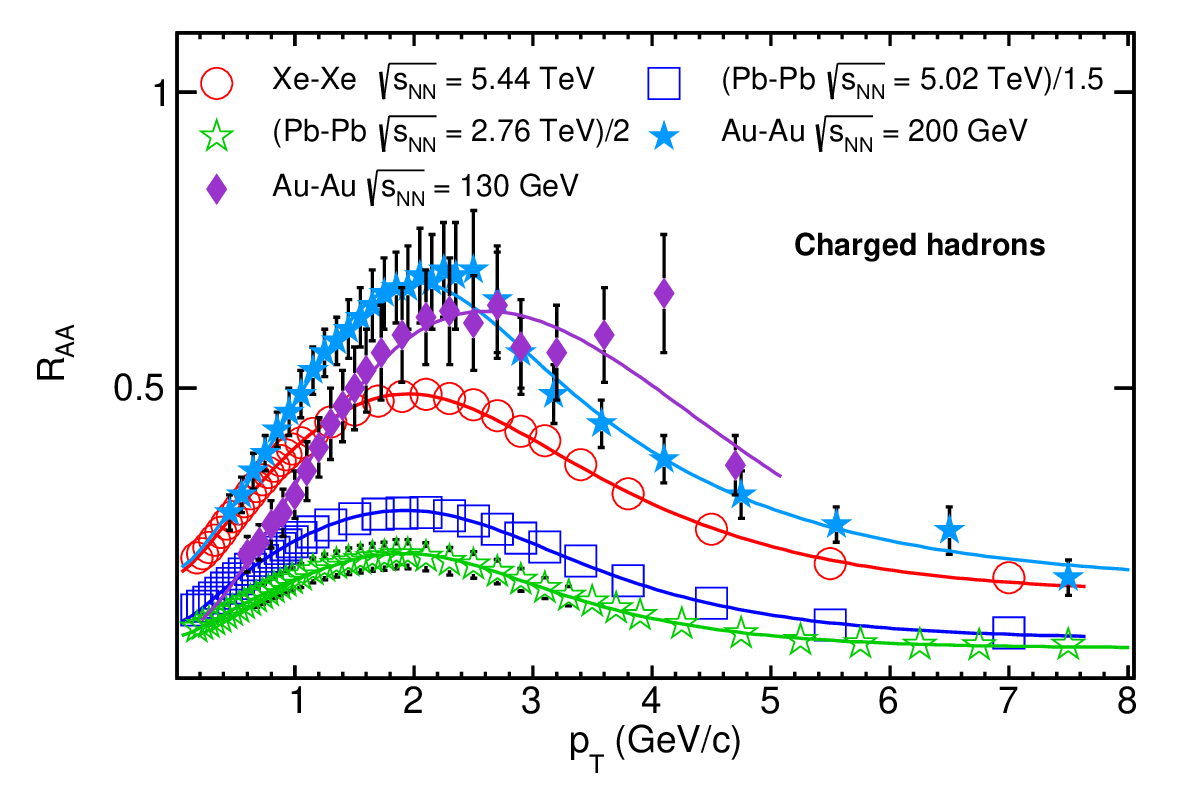}
  \caption{Nuclear modification factor of charged hadrons produced in most central $XeXe$ collision at $5.44$ TeV, $PbPb$ collision at $5.02~\&~2.76$ TeV and $AuAu$ collision at $200~\&~130$ GeV fitted with the equation for $R_{AA}$ given in Eq.~\ref{RAA_final}. Some of the data points are scaled for better visibility.}
  \label{fig:RAA_charged}
  \end{figure}
  \begin{figure}[!h]
\centering
  \includegraphics[width=\ims\textwidth]{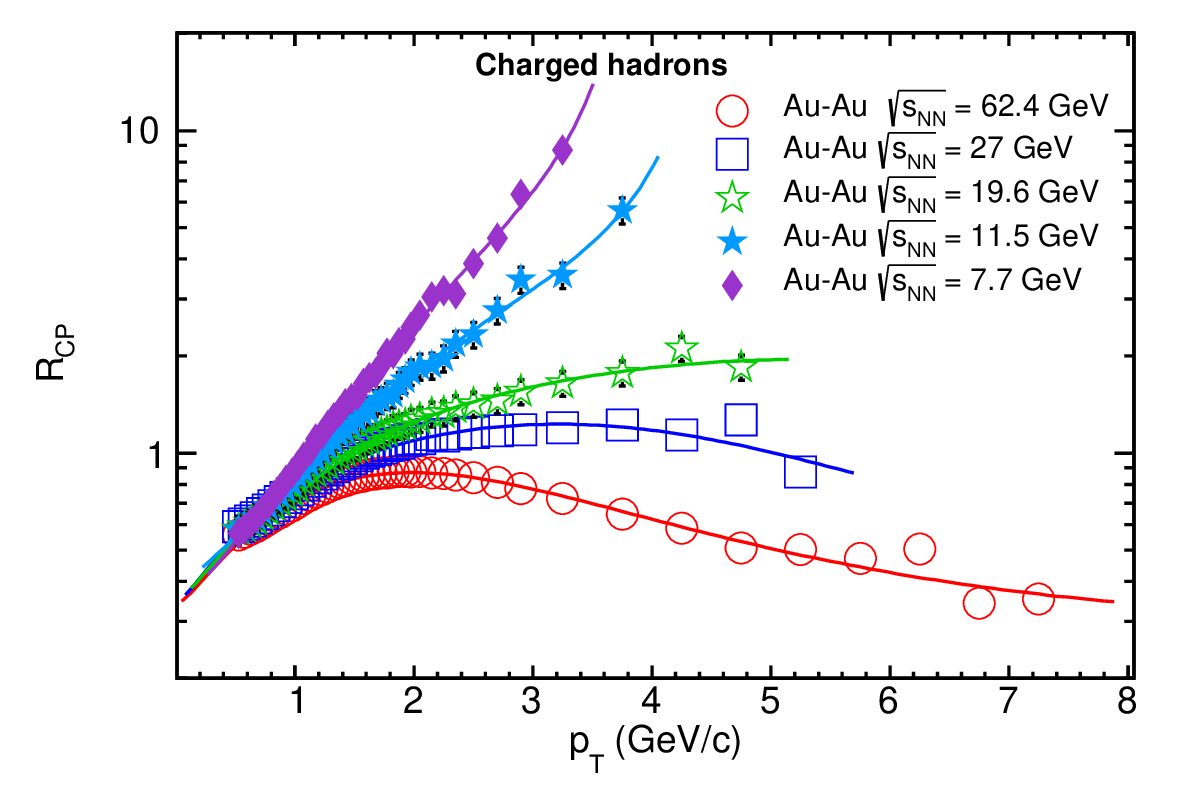}
  \caption{Nuclear modification factor ($R_{CP}$) of charged hadrons produced in most central $AuAu$ collision at $62.4,~27,~19.6,~11.5~\&~7.7$ GeV fitted with the equation for $R_{AA}$ given in Eq.~\ref{RAA_final}.}
  \label{fig:RCP_charged}
  \end{figure}

    \begin{figure*}
       \centering
\begin{subfigure}[b]{0.45\textwidth}
\includegraphics[width=\textwidth]{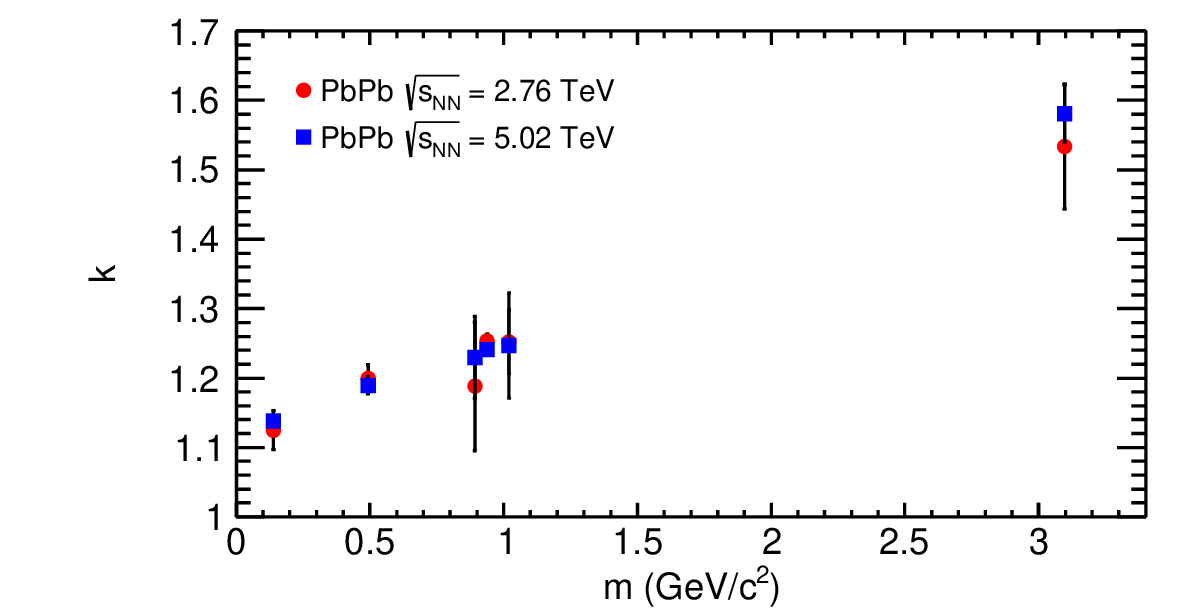}
\caption{Variation of parameter $k$ with mass of identified hadrons}
\label{fig:k_vs_mass}
\end{subfigure}
\begin{subfigure}[b]{0.45\textwidth}
\includegraphics[width=\textwidth]{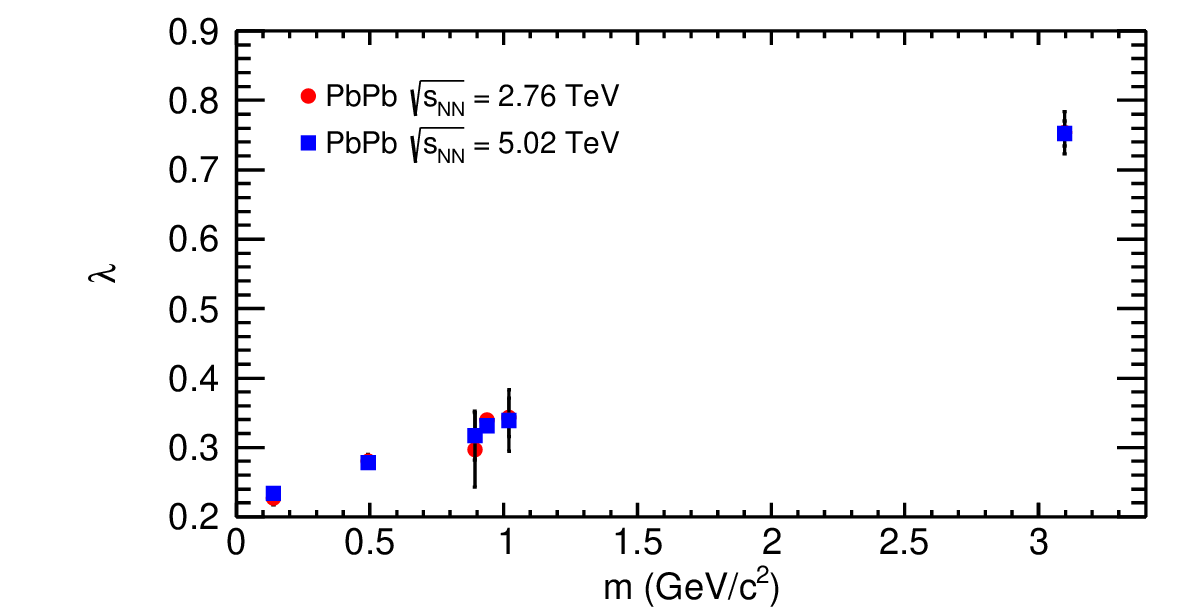}
\caption{Variation of parameter $\lambda$ with mass of identified hadrons}
\label{fig:lambda_vs_mass}
\end{subfigure}
\begin{subfigure}[b]{0.45\textwidth}
\includegraphics[width=\textwidth]{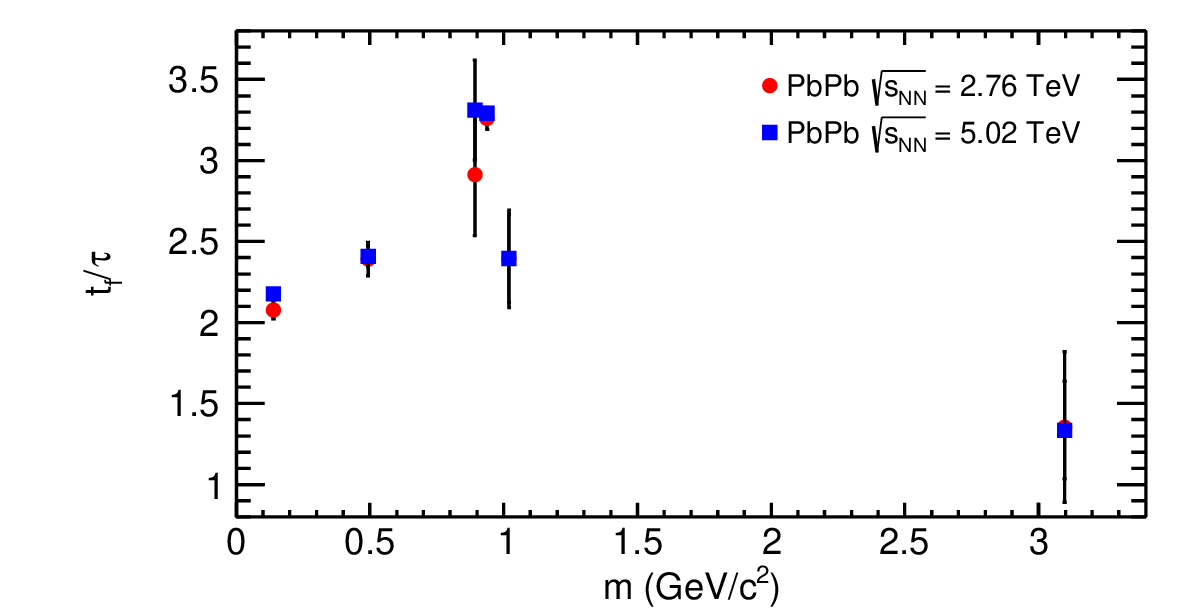}
\caption{Variation of parameter $t_f/\tau$ with mass of identified hadrons}
\label{fig:tf_tau_vs_mass}
\end{subfigure}

\caption{Variation of different fit parameters obtained by fitting the experimental data of identified hadrons produced in $PbPb$ collision at $2.76~\&~5.02$ TeV fitted with the equation for $R_{AA}$ given in Eq.~\ref{RAA_final}}
\label{fig:parameter_plots}
    \end{figure*}
    \section{Results and Discussion}
    Considering heavy-ion collision as incoherent superposition of $pp$ collisions, we should ideally get a value of $1$ across all $p_T$ ranges with some random variation because of the statistical and systematic uncertainties. However, looking into the experimental data of $R_{AA}$ in the Figs.~\ref{fig:RAA_identified_data}-\ref{fig:RAA_charged}, we observe a peculiar trend with $R_{AA}$ increasing in low $p_T$ range, saturating at a point followed by decrease as we move toward high $p_T$. From Fig.~\ref{fig:RAA_identified_data} we also observe that the $R_{AA}$ values reaches a minima around $6-7$ GeV/c and start to rise again. Initial rise in the data can be attributed to the Cronin enhancement, where multiple scattering of partons in heavy-ion causes a transverse boost and hence leading to a gain in transverse momentum of hadrons. The interplay of Cronin enhancement, gluon saturation \cite{Baier:2003hr, Jalilian-Marian:2003rmm} and jet quenching causes a maxima in intermediate $p_T$ range. This maxima is followed by a decline because of the suppression in the production of high $p_T$ hadron in heavy-ion collision due to dominance of jet quenching beyond $2-3$ GeV/c. At higher $p_T$, the $R_{AA}$ start to rise again because the suppression in the data of transverse momentum spectra start to level out around $8-10$ GeV/c as can be seen from the plot of experimental data of $p_T$ spectra (Fig. 1 in Ref.\cite{ALICE:2012aqc}), where the heavy-ion collision data is matched with $<T_{AA}>$ scaled $pp$ collision data of transverse momentum spectra.

Studying this peculiar pattern of $R_{AA}$ and its dependence on mass of particle species is of utmost importance in order to have a better understanding of the underlying physics. In our analysis, the upper limit of the $p_T$ is fixed to $8$ GeV/c because the impact of suppression on $p_T$ spectra starts to die out beyond this range. Also, for $p_T>8-10$ GeV/c the mass dependence of $R_{AA}$ data starts to disappear and we observe equal suppression for all particle species \cite{ALICE:2017ban, ALICE:2014juv}.

We begin our analysis by fitting the experimental data for $R_{AA}$ with the derived equation for nuclear modification factor Eq.~\ref{RAA_final}. The analysis is performed using the data analysis framework ROOT \cite{Brun:1997pa} and we have used the MINUIT \cite{James:1975dr} class for fitting the experimental data. The goodness of fit is tested using the $\chi^2/NDF$ values. While fitting the data using Eq.~\ref{RAA_final}, parameters $D'$, $k$, $\lambda$, $t_f/\tau$ $\&$ $q$ are considered as the fit parameters and the equilibrium temperature $T$ is fixed to the value of $160$ MeV throughout the analysis. For this work, we have considered the experimental data of charged hadron $R_{AA}$ at different RHIC and LHC energies including $AuAu$ collision at $130$ \cite{PHENIX:2002diz} $\&$ $200$ GeV \cite{STAR:2003fka}, $PbPb$ collision at $2.76$ \cite{ALICE:2012aqc} $\&$ $5.02$ TeV \cite{ALICE:2018vuu} and $XeXe$ collision at $5.44$ TeV \cite{ALICE:2018hza}. We have also analyzed the data of $R_{CP}$ measured at various energies ($62.4,~27,~19.6,~11.5~\&~7.7$ GeV \cite{STAR:2017ieb}) by STAR experiment as a part of Beam Energy Scan (BES) program. In order to study the dependence of fit parameters on mass of final state particles, the experimental data of $R_{AA}$ of the identified hadrons  ($\pi^++\pi^-,~K^++K^-,~p+\overline{p},~ K^{*0},~\phi~\&~J/\psi$) produced in $PbPb$ collision at $2.76$ TeV \cite{ALICE:2014juv, ALICE:2017ban, ALICE:2013osk} and $5.02$ TeV \cite{ALICE:2019hno, ALICE:2021ptz,  ALICE:2016flj} are also studied using the equation for $R_{AA}$ given in Eq.~\ref{RAA_final}. 

The fit results for identified hadrons obtained after fitting the experimental data with Eq.~\ref{RAA_final} is given in Figs.~\ref{fig:identified}~$\&$~\ref{fig:RAA_jpsi} for $2.76~\&~5.02$ TeV. Experimental data for charged pions, kaons and protons for two different energies are fitted in Figs.~\ref{fig:2760_1}~\&~\ref{fig:5020_1} and for neutral mesons $K^{*0}$ and $\phi$ in Figs.~\ref{fig:2760_2}~\&~\ref{fig:5020_2}. The corresponding fit for heavier charmonium bound state $J/\psi$ is provided in Fig.~\ref{fig:RAA_jpsi}. Further, we have also fitted the nuclear modification factor  of charged hadrons at various RHIC and LHC energies and the fit results are provided in Figs.\ref{fig:RAA_charged}~\&~\ref{fig:RCP_charged} for $R_{AA}~\&~R_{CP}$ respectively. From above Figs.\ref{fig:identified}-\ref{fig:RCP_charged}, we observe a good agreement between experimental data and theoretical formalism for $R_{AA}$ using BTE in RTA and considering $q$-Weibull as our final distribution function. The goodness of fit can also be verified by observing the $\chi^2/NDF$ values given in table 1. 

Apart from fitting the experimental data using above formalism, we can also extract the numerical values of fit parameters and use it to study whether there is some mass dependence in the suppression of particle production in heavy-ion collision. Since the identified particles considered in this study are made up of combination of different quark flavors ($\pi^+(u\overline{d},~K^+(u\overline{s}),~p(uud),~K^{*0}(d\overline{s}),~\phi(s\overline{s})~\&~J/\psi(c\overline{c}))$) and interaction between energetic partons and QGP medium is sensitive to the mass and color charge of partons, we should observe some difference in pattern of $R_{AA}$ depending on the mass of particle species. This difference in pattern will be encoded in the fit function and hence on the fit parameters. With an aim to explore the mass dependence, we have studied the variation of different fit parameter with the mass of particle species and the corresponding plots is provided in Fig.~\ref{fig:parameter_plots}. From Figs.~\ref{fig:k_vs_mass}~\&~\ref{fig:lambda_vs_mass}, we observe a linear mass dependence of parameters $k$ and $\lambda$ with their values increasing linearly with the increase in mass of the final state particle. 

As discussed in the Ref.~\cite{Dash:2018qln, Gupta:2023ggq}, parameter $k$ is linked to the onset of hard scattering processes and the value of this parameter increases with the rise in hardness of the spectra. Hardness of transverse momentum spectra refers to more number of particles in high $p_T$ region. Since jet quenching tends to reduce the number of high $p_T$ particles owning to partonic energy loss in the QGP medium, larger suppression should be related to softer (lesser hardness) spectra. Further, since the energy loss of energetic partons as they traverse through the QGP medium depends on their mass and color charge \cite{Sahoo:2025abz}, gluons will lose more energy than quarks. Also among the quarks of different flavor, lighter quark will lose more energy compared to heavier quarks because of the dead-cone effect. The QCD theory predicts that the gluon radiation from heavy quarks will be suppressed for smaller emission angles and for larger emission angle the radiation will be identical to that of lighter quarks \cite{Kluth:2023umf}. This suppression of gloun radiation is popularly known as the dead-cone effect and is observed by ALICE experiment at LHC \cite{ALICE:2021aqk}. Also, the radial flow in heavy-ion collision tends to make heavier particle gain more momentum compared to lighter particle \cite{Schnedermann:1993ws, Heinz:2013th} thereby pushing heavier particle toward higher $p_T$ and countering the impact of jet quenching. From above two arguments related to mass dependence of partonic energy loss and the mass ordering due to radial flow, we can conclude that the impact of jet quenching will be less in heavier particles making their spectra harder compared to lighter particles and hence leads to an increase in the value of parameter $k$ as observed in Fig.~\ref{fig:k_vs_mass}. 

Another important parameter whose mass dependence is studied in this work is the ratio of freeze-out time to relaxation time ($t_f/\tau$). Here freeze-out time refers to a time when interaction between particles ceases to exist and the relaxation time is the time which a non-equilibrated system takes to reach equilibrium. The study of this ratio is crucial in understanding the time evolution of a non-equilibrium system. A very low value ($t_f/\tau~<<~1$) signifies that freeze-out occur at a very early stage and the system was still far from equilibrium whereas a very large value ($t_f/\tau~>>~1$) tells us that the system had enough time to reach the equilibrium before the interaction among particles ceases to exist. We have studied the variation of $t_f/\tau$ with mass of particle and the result is presented in Fig.~\ref{fig:tf_tau_vs_mass}. The numerical values are close to unity  suggesting a near equilibrium scenario. We observe a clear mass dependence with values initially going up for lighter hadrons and then going down for hadrons made up of heavy quark flavors. This mass dependence support a mass differential freeze-out instead of a single freeze-out scenario. Heavier hadrons generally decouple earlier in time compared to lighter one as is evident from the study of freeze-out temperature in Ref.\cite{Lao:2015zgd, Waqas:2022fnl}. Also, relaxation time is higher for heavier particles because of their lower cross section \cite{Tripathy:2016hlg}. So, intuitively, we should get a decreasing trend with increase in mass and this is what we observed in Fig.~\ref{fig:tf_tau_vs_mass} for high mass hadrons. However, for hadrons made up of light quark flavors, we observe an increasing trend which does not go inline with the intuitive explanation. More studies in the direction of mass dependence of the ratio $t_f/\tau$ is required to have a better understanding of the equilibrium dynamics of system created in high energy collision.
\section{Conclusion}
In this work we have developed a new formalism to study the nuclear modification factor using the Boltzmann transport equation in relaxation time approximation. The novelty of this work lies in the choice of equilibrium $(f_{eq})$ and final state $(f_{fin})$ distribution function. For equilibrium distribution we have considered Boltzmann distribution which is the most natural choice for an equilibrium system and for final state distribution  we chose $q$-Weibull distribution. Important reasoning regarding the choice of this particular distribution function for final state $f_{fin}$ comes out from the results in Fig.~\ref{fig:tf_tau_vs_mass} where we observe that the values of $t_f/\tau$ is close to $1$. This value suggest a near equilibrium scenario so we should consider a distribution function which include some deviation from thermal equilibrium measured by non-extensivity parameter $q$. Further, quenching effect are most pronounced in intermediate $p_T$ range, hence our distribution function should have applicability over a broader $p_T$ range unlike Tsallis, BW or TBW which deviates beyond $3-4$ GeV/c $p_T$ values. These two set of reasoning justifies the choice of $q$-Weibull distribution function which includes non-extensivity parameter $q$ and nicely explain transverse momentum spectra over broader $p_T$ range \cite{Dash:2018qln, Gupta:2023ggq}.

Using this formalism we have studied the experimental data of nuclear modification factor of the charged hadrons produced in different collision system such as $AuAu$, $PbPb$, $\&$ $XeXe$ and measured at various RHIC and LHC energies ranging from $7.7$ GeV all the way upto $5.44$ TeV. We also analyzed the $R_{AA}$ of identified hadrons produced in $PbPb$ collision at $2.76$ TeV $\&$ $5.02$ TeV and observe a good agreement between model and experimental data across various energies and particle species. We also studied the dependence of different fit parameters on mass of final state particles and observe a linear mass dependence for parameter $k$ and $\lambda$. The trend in parameter $k$ can be attributed to the reduction in quenching effect and increase in hard scattered partons with mass. We also studied the variation of parameter $t_f/\tau$ with mass and observe a peculiar trend whose explanation require a detail theoretical study of the mass dependence of freeze-out and relaxation time.

In conclusion, we have developed a theoretical model that nicely explain the nuclear modification factor and provides us with a tool to study the pattern in experimental data of $R_{AA}$ across broad range of energies. 


%
%

%
%

\end{document}